\documentclass[a4paper,12pt]{article}
\usepackage{epsfig} 
\usepackage{hyperref}   
\usepackage{amsmath}
\usepackage{psfrag}
\usepackage{jpc}
\usepackage{overcite}  
 \usepackage{setspace}

\newcommand{\rem}[1]{}

\begin{document}

\title{ An optimized algebraic basis for molecular potentials.  }

\author{Andrea Bordoni\thanks{Corresponding author. E-mail: andrea.bordoni@unimi.it}
\ and Nicola Manini\\
{\it Dipartimento di Fisica, Universit\`a di Milano,}\\
{\it Via Celoria 16, 20133 Milano, Italy}
} 

\date{July 27 2007} 
\maketitle

\begin{abstract}

The computation of vibrational spectra of diatomic molecules through
the exact diagonalization of algebraically determined matrixes based
on powers of Morse coordinates is made substantially more efficient by
choosing a properly adapted quantum-mechanical basis, specifically
tuned to the molecular potential.  A substantial improvement is
achieved while still retaining the full advantage of the simplicity
and numerical light-weightedness of an algebraic approach. In the
scheme we propose, the basis is parameterized by two quantities which
can be adjusted to best suit the molecular potential through a simple
minimization procedure.

\end{abstract}


\section{Introduction}

In a previous work \cite{Bordoni_2006}, an algebraic method for the
computation of vibrational spectra of diatomic molecules was
introduced.  Although this is a 1-dimensional (1D) problem, thus an in
principle trivial task, the algebraic method shows substantial
advantages over both the real-space grid solution of the
Schr\"{o}dinger equation and harmonic-oscillator-based
techniques. These advantages are especially important for extensions
to the multidimensional problem of polyatomic vibrations.

The expansion of the molecular potential in powers of the
Morse-potential related quantity $v(x)=e^{-\alpha (x-x_0)}-1$, namely
\begin{equation}\label{Morse-Exp2}
V_d(x)= \sum_{k=2}^{N_{\rm max}} a_k \, \left(v(x)\right)^k
,
\end{equation}
allows an efficient and accurate approximation of a well-behaved
molecular potential in the whole energy range, from the minimum region
to the dissociation threshold, generally involving a moderate number
$N_{\rm max}+1$ parameters $a_2,\dots,\alpha$ and $x_0$.  Even potentials
substantially distorted with respect to the Morse potential can be
treated successfully.  With the potential expressed in the form of
Eq.~(\ref{Morse-Exp2}), the complete Hamiltonian
\begin{equation}\label{Ham_exp2}
\hat{H} \equiv
-\frac{\hat{p}_x^2}{2 \mu} + V_d(x) 
\end{equation}
(here $\mu$ is the reduced mass of the 2-body problem and $x$ is the
radial coordinate) can be represented on a quantum-mechanical basis of
choice.

The accuracy and efficiency of the direct diagonalization methods rely
both on the accuracy of the potential approximation of
Eq.~(\ref{Morse-Exp2}) and on the properties of the selected basis.
The basis had better be complete but also manageable, i.e.\ related to
the algebraic properties of $v(x)$, so that the evaluation of the
matrix elements can be done rapidly and without approximations: this
will be needed especially in view of extensions to polyatomic
molecules.

\section{The Basis}

Previous research \cite{Bordoni_2006,Ben-Mol-Al,su11} showed that the
 basis
\begin{equation}\label{QNSB2}
\phi_n(y)=\sqrt{\frac{\alpha n!}{\Gamma (2\sigma +n)}}
\,y^{\sigma} e^{-\frac{y}{2}}L_n^{2\sigma -1}(y),~\sigma>0,\quad
 n=0,~1,~2,\dots\,,
\end{equation}
with
\begin{equation}\label{ydef2}
y(x)=(2s+1)\, e^{-\alpha(x-x_0)}\,,
\end{equation}
can be usefully employed in general diatomic contexts, with the
special choice
\begin{equation}\label{SigFix2}
\sigma=s-[s] \,,
\end{equation}
where $[s]$ indicates the integer part of $s$, and with $s$ related to
the Morse term $a_2 \,(v(x))^2$ in the potential expansion
(\ref{Morse-Exp2}), by
\begin{equation}\label{SFix2} 
s=\frac{\sqrt{2 \mu a_2}}{\hbar \alpha}-\frac{1}{2}\,.
\end{equation}
With the conditions (\ref{SigFix2},\ref{SFix2}) the basis
(\ref{QNSB2}) was named {\it quasi number state basis} (QNSB)
\cite{Ben-Mol-Al}.  In the present work, we only assume $\alpha$ and
$x_0$ in Eqs.~(\ref{QNSB2},\ref{ydef2}) are the same as in the
potential expansion (\ref{Morse-Exp2}), and that $\sigma >0$ and $s>
-\frac{1}{2}$, but release all additional unnecessary conditions on
$\sigma$ and $s$, for example those expressed by
Eqs.~(\ref{SigFix2},\ref{SFix2}), or the condition defined by Tennyson
and Sutcliffe\cite{Tennyson82,Tennyson04} (TS):
\begin{equation}\label{SutTennCond}
\sigma=\frac{[2 s]+2}{2}\,,
\end{equation}
with $s$ fixed by Eq.~(\ref{SFix2}). Equation~(\ref{QNSB2}) thus
defines a $(s,\sigma)$-parameterized family of bases, {\it
generalized} QNSB (GQNSB), all sharing the following main features:
(i) the basis (\ref{QNSB2}) is complete;
 (ii) the kinetic and potential operators can be written in terms of
generalized ladder operator as specified below, so that (iii) the
matrix elements of a vast class of relevant operators is computable
easily and exactly by means of simple algebraic relations
\cite{Bordoni_2006}.

Even though all infinite GQNSB's are substantially equivalent,
regardless of $s$ and $\sigma$, different bases characterized by
different values of $s$ and $\sigma$ show different performances when
truncated to a finite number $N_s$ of states and applied to a given
quantum mechanical problem specified by $\mu,\alpha,a_2,a_3, \dots,
a_{N_{\rm max}}$. Indeed, the purpose of the present work is to
demonstrate that a properly chosen truncated GQNSB can improve the
efficiency of the computation substantially, compared to earlier
choices \cite{Bordoni_2006,Tennyson82}.

\section{Matrix elements}

We follow here the same approach \cite{Bordoni_2006} derived from SUSY quantum mechanics
\cite{Ben-Mol-Al,SusyRev}.
We introduce the generalized Morse ladder operators \cite{Bordoni_2006,Ben-Mol-Al}
\begin{eqnarray}\label{SUSYGenLadd2}
\hat{A}(q)&=&
q\hat{I} - \frac{\hat{y}}{2} + \frac{i}{\hbar \alpha}\hat{p}_x\,\\ \nonumber
\hat{A}^{\dagger}(q)&=&
q\hat{I} - \frac{\hat{y}}{2}  - \frac{i}{\hbar \alpha}\hat{p}_x \ ,
\end{eqnarray}
parameterized by the real quantity $q$\cite{Nota1}.
 These operators, with a suitable choice of $q$, act
on the states (\ref{QNSB2}) of the GQNSB as ladder operators:
\begin{eqnarray}\label{AAdagComm}
\hat{A}(\sigma+n) \,\phi_n &=&  C_n \,\phi_{n-1}\\ \nonumber
\hat{A}^{\dagger}(\sigma+n) \,\phi_n &=&  C_{n+1} \,\phi_{n+1} \,,
\end{eqnarray}
where
\begin{equation}\label{CCdef2} 
C_{n}=\sqrt{n(n+2\sigma-1)}
\,.
\end{equation} 

According to Eqs.~(\ref{SUSYGenLadd2},\ref{AAdagComm}), $\sigma$ links
the parameterized basis (\ref{QNSB2}) to the corresponding family of
generalized ladder operators. Thus, each and every basis of the form
of Eq.~(\ref{QNSB2}) can be managed algebraically in this formalism,
for any given choice of $\sigma>0$.  In practice, the eigenfunctions
(\ref{QNSB2}) depend explicitly on $s,\alpha$ and $x_0$, beside
$\sigma$. We fix $x_0$ to the position of the minimum of the potential
(\ref{Morse-Exp2}), as it would not provide a substantial advantage
otherwise. Likewise, we select for $\alpha$ the same value as in the
potential expansion, because otherwise all relevant matrix
representations would be dense rather than sparse 
\cite{Nota2}%\footnote{The use of
.
With these constraints on $x_0$ and $\alpha$, an arbitrary $s$ can be
usefully employed in the basis definition: for any $s$ value, the
momentum operator $p_x$ and the multiplication operator
$e^{-\alpha(\hat{x}-x_0)}$ can be written in terms of the ladder
operators (\ref{SUSYGenLadd2}):
\begin{eqnarray}\label{Expdef2}
e^{-\alpha(\hat{x}-x_0)} &=& \frac{ 2q \hat{I} - \left[\hat{A}^{\dagger}(q)+\hat{A}(q)\right]}{(2 s +1)}\,,
\\
\hat{p}_x &=& \frac{\hbar \alpha}{2i} \left[\hat{A}(q)-\hat{A}^{\dagger}(q)\right] \,,\label{PXdef2q2}
\end{eqnarray}
where also the $\hat{A}$ operators depend implicitly on the $s$ parameter
appearing in the definition (\ref{ydef2}) of $\hat{y}$.  On the GQNSB
(\ref{QNSB2}), the matrix elements of any physical operator
expressed as a polynomial of $e^{-\alpha(\hat{x}-x_0)}$ and $p_x$ can
be computed algebraically since Eqs.~(\ref{Expdef2},\ref{PXdef2q2})
express them in terms of the ladder operators of the corresponding
specialized basis.
We derive here explicitly the algebraic form of the Morse Hamiltonian
for general $q$ and $s$.

Using Eq.~(\ref{PXdef2q2}), the kinetic operator
$\hat{K}=\frac{\hat{p}_x^2}{2 m}$ becomes
\begin{equation}\label{KinGen2}
\hat{K}=
-\frac{\hbar^2 \alpha^2}{8 m} [\hat{A}^2(q) + \hat{A}^{\dagger 2}(q) - \hat{A}(q) \hat{A}^{\dagger}(q) - \hat{A}^{\dagger}(q) \hat{A}(q)]\,.
\end{equation}
By applying the commutation relations 
\begin{eqnarray}\label{A-comm2}
[\hat{A}(q),\hat{A}^{\dagger}(q')] &=& (q+q')I -( \hat{A}(q)+ \hat{A}^{\dagger}(q')),\\ \nonumber
[\hat{A}(q),\hat{A}(q')] &=& [\hat{A}^{\dagger}(q),\hat{A}^{\dagger}(q')]=0\,,
\end{eqnarray}
$\hat{K}$ reduces to
\begin{equation}\label{KinGen}
\hat{K}=-\frac{\hbar^2 \alpha^2}{8 m} [\hat{A}^2(q) + \hat{A}^{\dagger 2}(q) - 2q \hat{I} +\hat{A}(q) +\hat{A}^{\dagger}(q) - 2 \hat{A}^{\dagger}(q) \hat{A}(q)]\,.
\end{equation}
Likewise, powers of $e^{-\alpha(\hat{x}-x_0)}$ appearing in the
potential-energy operator are obtained starting from
Eq.~(\ref{Expdef2}). For example,
\begin{eqnarray}\label{Exp-gen}
&&e^{-2\alpha(\hat{x}-x_0)} =\\ \nonumber
&=& \frac{1}{(2 s +1)^2} \{4 q^2 \hat{I} -4 q [\hat{A}^{\dagger}(q)+\hat{A}(q)] + \hat{A}^2(q) + \hat{A}^{\dagger 2}(q) +\hat{A}(q) \hat{A}^{\dagger}(q) + \hat{A}^{\dagger}(q) \hat{A}(q)\}\\ \nonumber
&=&\frac{1}{(2 s +1)^2} \{2 (2q^2 +q) \hat{I} -(4 q +1) [\hat{A}^{\dagger}(q)+\hat{A}(q)] + 2 \hat{A}^{\dagger}(q) \hat{A}(q) + \hat{A}^2(q) + \hat{A}^{\dagger 2}(q)\}\,.
\end{eqnarray}
Thus, the Morse-potential term reads
\begin{eqnarray}\label{PotentialGen2}\nonumber
(v(\hat{x}))^2&=&  \frac{1}{(2 s +1)^2} \{2 (2q^2 +q) \hat{I} -(4 q +1) [\hat{A}^{\dagger}(q)+\hat{A}(q)] + 2 \hat{A}^{\dagger}(q) \hat{A}(q) \\ \nonumber
&+& \hat{A}^2(q) + \hat{A}^{\dagger 2}(q)\} - \frac{2}{(2 s +1)} \{2q \hat{I} - [\hat{A}^{\dagger}(q)+\hat{A}(q)]\}\\ \nonumber
&=&\frac{1}{(2s+1)^2} \{ (4 q^2 -2q -8sq) \hat{I} + (4s-4q+1)  [\hat{A}^{\dagger}(q)+\hat{A}(q)] \\  
&+& 2 \hat{A}^{\dagger}(q) \hat{A}(q) 
+ \hat{A}^2(q) + \hat{A}^{\dagger 2}(q)\}\,.
\end{eqnarray}

Accordingly, the Morse Hamiltonian $\hat{H}_M=\hat{K} +a_2
(v(\hat{x}))^2$ is expressed in algebraic form as
\begin{eqnarray}\label{HamCompGen2}\nonumber
\hat{H}_M &=&  \left[ \frac{2 a_2}{(2s+1)^2} + \frac{\hbar^2 \alpha^2}{4 m}\right] \hat{A}^{\dagger}(q) \hat{A}(q)  
+ q \left[\frac{2 a_2}{(2s+1)^2} (2q -1-4s)  + \frac{\hbar^2 \alpha^2}{4 m}  \right] \hat{I}\\\nonumber
&+& \left[\frac{a_2}{(2s+1)^2} (4s -4q +1) -\frac{\hbar^2 \alpha^2}{8 m}\right] [\hat{A}^{\dagger}(q) +\hat{A}(q)]\\
&+& \left[\frac{a_2}{(2s+1)^2} -\frac{\hbar^2 \alpha^2}{8 m}\right] [\hat{A}^2(q) + \hat{A}^{\dagger 2}(q)]\,.
\end{eqnarray}
 The representation of Eq.~(\ref{HamCompGen2}) shows that the Morse
 Hamiltonian is generally 5-band diagonal on a GQNSB of the form
 (\ref{QNSB2}). We stress that the expression ({\ref{HamCompGen2})
 holds for \emph{any} choice of parameters $s$ and $q$, regardless of
 them being connected to any specific physical constraint.

If the condition 
\begin{equation}\label{TridiagCond2}
\frac{a_2}{(2s+1)^2} = \frac{\hbar^2 \alpha^2}{8 m}\,
\end{equation}
(equivalent to Eq.~(\ref{SFix2})) is satisfied, then the last term,
proportional to $[\hat{A}^2(q) + \hat{A}^{\dagger 2}(q)]$ drops from
$\hat{H}_M$.  In other words, the choice of the parameter $s$ of
Eq.~(\ref{SFix2}) makes the Morse Hamiltonian \emph{tridiagonal} on
the corresponding GQNSB basis, irrespective of $q$.  Under this
special condition (\ref{TridiagCond2}), the Morse Hamiltonian
simplifies to:
\begin{equation}\label{HamCompGenSfixed2}
\hat{H}_M= \frac{4 a_2}{(2s+1)^2}\left\{[\hat{A}^{\dagger}(q)+\hat{A}(q)]  \left(s -q\right) 
+ \hat{A}^{\dagger}(q) \hat{A}(q) +  (q^2 -2qs) \hat{I}\right\}.
\end{equation}
The form of Eq.~(\ref{HamCompGenSfixed2}), indicates that by further
setting
\begin{equation}\label{qSimp}
q=s\,,
\end{equation}
the operator form of the Hamiltonian simplifies even more, and the
Morse Hamiltonian factorizes as:
\begin{equation}\label{HamCompGenSfixedqeqs2}
\hat{H}_M = 4 \frac{a_2}{(2s+1)^2} [\hat{A}^{\dagger}(s) \hat{A}(s) - s^2  \hat{I}]\,,
\end{equation}
which recovers the algebraic form of the Morse Hamiltonian of
previous works\cite{Bordoni_2006,Ben-Mol-Al}.

The use of different values of $q$ and $s$ produces a GQNSB, where the
algebraic computation of the matrix elements of the Hamiltonian
(\ref{Ham_exp2}) is not significantly more intricate: in particular on
a GQNSB, the Morse Hamiltonian is 5-band diagonal, rather than
tridiagonal \cite{Nota3}, and
higher powers of $(v(x))^k$ in Eq.~(\ref{Morse-Exp2}) generate
$(2k+1)$-band diagonal matrices (like in the QNSB).

For practical potentials, usually substantially distorted from the
 pure-Morse $(v(x))^2$ term, the actual eigenfunctions can be
 represented poorly by the $[s]+1$ Morse bound states, or equivalently
 by their QNSB counterparts: to achieve a good convergency of all
 eigenfunctions, the QNSB often needs to be complemented by a large
 number of states, far beyond $[s]+1$.
A suitably chosen GQNSB can thus prove significantly more efficient,
especially in a multi-oscillator polyatomic context.

\section{GQNSB parametric dependency}
\label{BasisDepParam}

\rem{%Fig 1
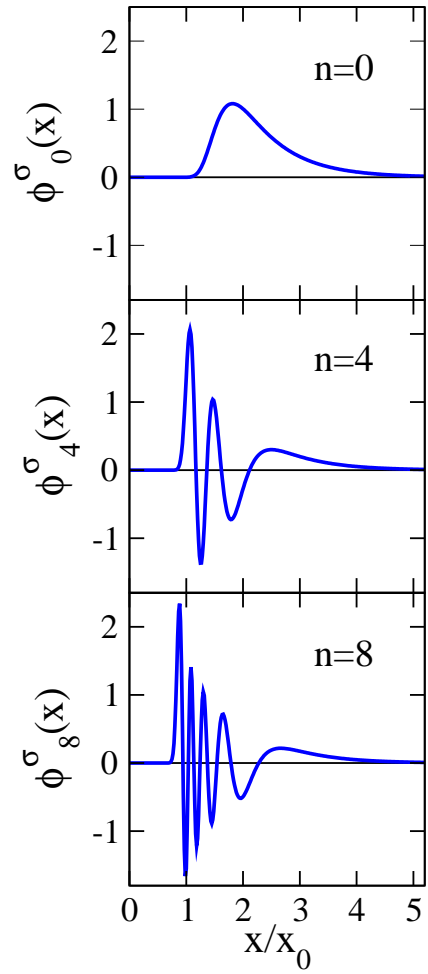
\begin{figure}
\centerline{ 
\begin{tabular}{c}
\epsfig{file=xQNSBCombD.eps,width=5.5cm,clip=}\\
\end{tabular}}\caption{QNSB  wavefunctions, for $n=0$, $n=4$ and $n=8$, compatible with a Morse problem characterized by $x_0=1,\alpha=4/x_0,a_2=625$, in units where $\mu=1, \hbar=1$, so that $s=8.34, \sigma=0.34$.  \label{QNSBEigenGraf2}}
\end{figure}    
}% Fine rem Fig 1

The shape of the wavefunctions (\ref{QNSB2}) depends on the four
parameters $x_{0}, \alpha, s$ and $\sigma$: different shapes imply
different convergence properties when employed to build the matrix
representation of the Hamiltonian. A brief analysis of the dependency
of the shape of GQNSB states on the various parameters can be useful
to gain some insight in their role.  Figure \ref{QNSBEigenGraf2} shows
the profile of three states of the form (\ref{QNSB2}), under
conditions (\ref{SigFix2}) and (\ref{SFix2}).  Note that the $n=0$
state is located substantially at the right of the Morse equilibrium
position $x_0$, and that further states move in toward $x_0$ for
increasing $n$.  This contrasts with the behavior of a basis of energy
eigenstates of a well centered in $x_0$.
\rem{ %Fig 2
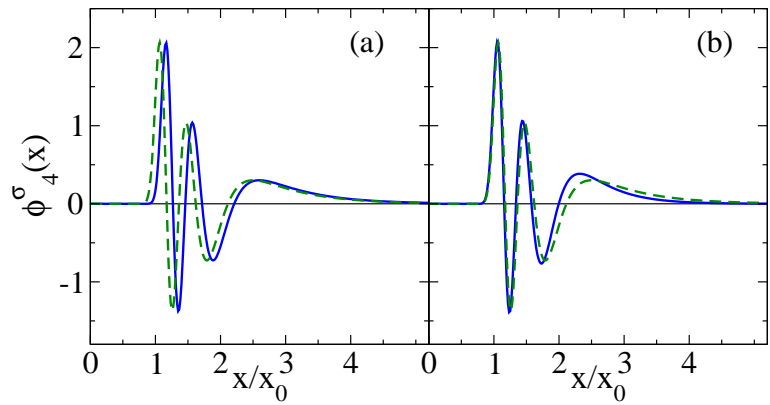
\begin{figure}  
\centerline{\begin{tabular}{c}
\epsfig{file=xQNSBMidCombD20.5b.eps,width=10.0cm,clip=}\\
\end{tabular}}\caption{Variation of the $n=4$ GQNSB wavefunction for a $50\%$ increase in the parameters $s$ (a), or $\sigma$ (b), solid line, with respect to the QNSB starting wavefunction (dashed line) corresponding to $s=8.34$, $\sigma=0.34$, $x_0=1$, $\alpha=4/x_0$ like in Fig.~\ref{QNSBEigenGraf2}.  \label{QNSB+varparamEigenGraf2}}  
\end{figure} 
} % Fine Rem Fig 2
Figure~\ref{QNSB+varparamEigenGraf2} illustrates the behavior of a
GQNSB wavefunction, Eq.~(\ref{QNSB2}), after variation of the
parameters $s$ and $\sigma$ involved relative to the QNSB values,
Eqs.~(\ref{SigFix2}, \ref{SFix2}).
The dependence on the $s$ parameter
(Fig.~\ref{QNSB+varparamEigenGraf2}a) is weak: by increasing $s$, the
eigenfunction shifts almost rigidly towards the outer region.
The $\sigma$-dependence (Fig.~\ref{QNSB+varparamEigenGraf2}b) is less
trivial: for larger $\sigma$, the wavefunction deforms and shrinks,
concentrating toward the region of the minimum, and decaying more
rapidly at large $x$.  The role of the $\sigma$ parameter is
particularly important: as the $n^{\rm th}$ GQNSB wavefunction
(\ref{QNSB2}) has the general form
\begin{equation}\label{StatoQNSB2} 
\phi_n(y) \propto e^{-y/2}y^{\sigma} Pol[y,n]\,,
\end{equation}
($Pol[y,n]$ stands for a polynomial of degree $n$ in the variable
$y$), $\sigma$ controls the {\it decay rate of the wavefunctions} for
$y\rightarrow 0$, i.e.\ at the dissociation region.
In particular, by choosing small $\sigma$, the basis wavefunctions
spread away from the well region thus improving the convergency of
high-energy states, possibly at the expense of quality of the
low-energy states in the well.
Equation~(\ref{StatoQNSB2}) and Fig.~\ref{QNSB+varparamEigenGraf2}
show that the general shape and in particular the amount of
localization of the GQNSB wavefunctions can be tuned freely by
choosing suitable $s$ and $\sigma$ parameters: this allows improving
the variational efficiency of a truncated GQNSB for a specific
quantum-mechanical problem.

\section{Optimization of the basis parameters}
\label{ParamOptim}

\rem{ % Fig 3
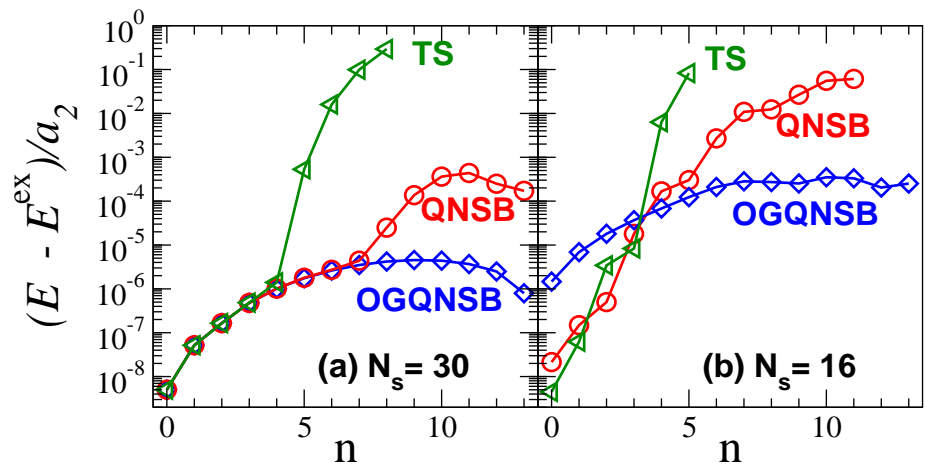
\begin{figure}\centerline{
\begin{tabular}{c}
\epsfig{file=Quart1_30_16scart2.eps,width=12cm,clip=}
\end{tabular}} 
\caption{Discrepancies $(E_n-E^{ex}_n)/a_2$ of the individual
eigenvalues $n$ for the potential $V(x)= a_2 [(v(x))^{2} +
(v(x))^{4}]$. (a) $N_s=30$ (OGQNSB with $s_{\rm min}=20.01$,
$\sigma_{\rm min}=0.435$); (b) $N_s=16$ (OGQNSB with $s=14.47$,
$\sigma=0.314$), for all bound states. The OGQNSB (diamonds)
discrepancies are compared to those based on the QNSB (circles), and
to the GQNSB based on the choice of $(s,\sigma)$ made by Tennyson and
Sutcliffe \cite{Tennyson82,Tennyson04} (triangles).\label{DiffComp2} }
\end{figure}  
} %Fine rem Fig 3
 
Assume that the exact $N_b$ bound state eigenvalues $E^{\rm ex}_i$ of the
Hamiltonian are known; we can measure the RMS discrepancy of the
discrete spectrum due to basis-incompleteness by
\begin{equation}\label{QNSB+Minim2}
\tilde{\Delta}^2=\frac{1}{N_b}\sum_{n=0}^{N_b-1} (E_n-E^{\rm ex}_n)^2\,,
\end{equation}
in terms of the numerical eigenvalues $E_i$, obtained by diagonalizing
the matrix of $\hat{H}$, Eq.~(\ref{Ham_exp2}), on a finite GQNSB
composed by the first $N_s~(>N_b)$ states and parameterized by $s$ and
$\sigma$.  For fixed $N_s$ we can search for the optimal $s_{\rm min}$ and
$\sigma_{\rm min}$ that make $\tilde{\Delta}$ minimum.

In fact, the {\it a priori} knowledge of the exact eigenvalues
$E^{\rm ex}_i$ is not necessary: due to the variational nature of basis
truncation, a ``better'' basis makes all eigenvalues $E_i$
lower. Accordingly, the optimal $s_{\rm min}$ and $\sigma_{\rm min}$
parameters can be defined as those producing the lowest eigenvalue
spectrum for the assigned basis size $N_s$, i.e. those minimizing
\begin{equation}\label{QNSB+MinimEig2}
\Delta=\frac{1}{N_b}\sum_{n=0}^{N_b-1} E_n\,.
\end{equation}
  This approach only requires that the number $N_b$ of bound
eigenstates is known.  Of course, the
number $N_b$ of bound states can be determined once and for all, for
example by means of a calculation on a very extended QNSB.
The minimization of $\tilde{\Delta}$ and of $\Delta$ leads generally to
slightly different results, but the following qualitative discussion
applies equally well to both schemes. Unless specified, for the
determination of $s_{\rm min}$ and $\sigma_{\rm min}$, we minimize $\Delta$ as
defined in Eq.~(\ref{QNSB+MinimEig2}), and compare $\Delta$ to its
fully-converged value $\Delta_0$ computed on a largely complete basis.

In a typical application of the GQNSB, one starts from a molecular
potential energy expressed in terms of an expansion of the form of
Eq.~(\ref{Morse-Exp2}).
Before considering realistic dimers (H$_2$ and Ar$_2$), we illustrate
the properties of the optimized GQNSB for a simple toy potential
defined by
\begin{equation}\label{ToyModel}
N_{\rm max}=4,\,a_2=a_4=625,\,a_3=0,\, \alpha=4\,,\,{\rm and}\,\, x_0=1,
\end {equation}
which we solve combined with a kinetic term specified by
$\hbar=1,~\mu=1$.
We minimize $\Delta$ with respect to $s$ and $\sigma$, for two fixed
numbers of basis states $N_s=30$ and $16$. Figure \ref{DiffComp2}(a)
shows the values of the individual eigenvalue discrepancy $(E_{n}
-E_{n}^{\rm ex})/a_2$ for the potential (\ref{ToyModel}), for the
QNSB, for an optimized GQNSB (OGQNSB), and for $s$ and $\sigma$ chosen
according to the prescription of TS \cite{Tennyson82,Tennyson04}. The
optimized parameters of the $N_s=30$ OGQNSB are $s_{\rm min}=20.01$ and
$\sigma_{\rm min}=0.435$, to be compared with the QNSB ones $s=8.338$ and
$\sigma=0.338$, and those chosen according to the prescription of
TS~\cite{Tennyson82,Tennyson04} $s=8.338$
and $\sigma=9$.  For this potential $\Delta_0=-444.90$, and the
corresponding $\Delta-\Delta_0$ are $3\cdot 10^{-6}$ for the OGQNSB
($\Delta$ equaling $\Delta_0$ to 5 decimal digits), $0.061$ for the
QNSB, and $457$ for the TS choice.  Both QNSB and the OGQNSB retrieve
all the bound states, but the OGQNSB produces much better converged
eigenenergies, especially near dissociation. The TS basis instead
yields only $9$ of the $14$ bound states, only few of which are
converged within $10^{-2} \,a_2$, which explains the large discrepancy
$\Delta-\Delta_0$.
\rem{ % Fig 4
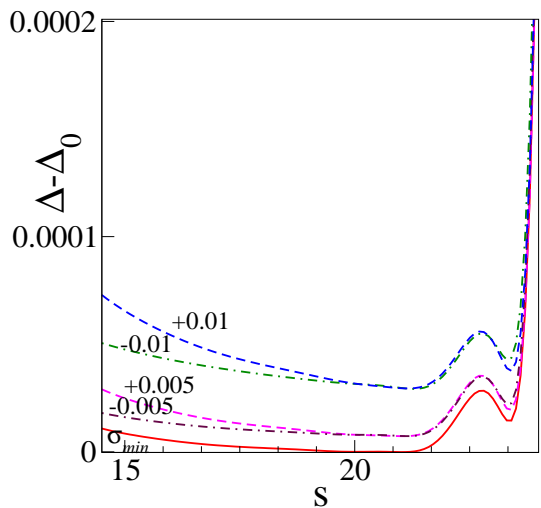
\begin{figure}
\centerline{
\begin{tabular}{c}
\epsfig{file=EsseperdeltasigI2newb.eps,width=7cm,clip=}\\
\end{tabular}}\caption{$s$ dependence of $\Delta$ Eq.~(\ref{QNSB+MinimEig2}), for $V(x)$ defined in Eq.~(\ref{ToyModel}), computed with the GQNSB of $N_s=30$ elements, as a function of $s$, and for $\sigma$ fixed to $\sigma_{\rm min}$, to $\sigma_{\rm min} \pm0.005,$ and to $\sigma_{\rm min} \pm 0.01$. \label{sandsigmadependence2}}
\end{figure}
} % Fine rem Fig 4  
 
Figure \ref{DiffComp2}(b) shows the same individual discrepancies
obtained with a basis of $N_s=16$ states instead of $30$.  The $s$ and
$\sigma$ values of the QNSB and the TS basis are of course unchanged,
while for the OGQNSB they change to $s_{\rm min}=14.47$ and
$\sigma_{\rm min}=0.314$.  The discrepancies $\Delta-\Delta_0$ deteriorate
to $0.106$, $87.35$, and $4045.9$ for OGQNSB, QNSB and TS
respectively.  Clearly the OGQNSB maintains a fair accuracy throughout
the spectrum, by allowing for slightly less accurate lowest bound
states, at the benefit of those near dissociation. In contrast, the
$N_s=16$ QNSB fails in obtaining the two bound states closest to
dissociation, and the TS basis only produces $6$ bound states.  Thus,
basis parameters optimization allows a substantial improvement of the
accuracy of the results, with the same computational cost. In other
words, the convergence speed of the computation can be improved
drastically by means of a suitable choice of $s$ and $\sigma$, for
example Fig.~\ref{DiffComp2} demonstrates an equal accuracy of the
OGQNSB of $16$ states and the QNSB of $30$ states.

Figure \ref{sandsigmadependence2} illustrates a typical $s$ dependence
of the total discrepancy $\Delta-\Delta_0$: for $\sigma$ equal to its
optimal value $\sigma_{\rm min}$ (solid curve), as $s$ approaches the
optimal $s_{\rm min}$ value from below, $\Delta$ decreases relatively
slowly, while for $s$ increasing beyond $s_{\rm min}$, $\Delta$ grows very
steeply.  The $\sigma$ dependence of $\Delta$ has a sharp and roughly
symmetrical deep minimum around $\sigma_{\rm min}$.

The reason for the observed $s$ and $\sigma$ dependencies of $\Delta$
is related to the GQNSB wavefunction profiles of
Figs.~\ref{QNSBEigenGraf2} and \ref{QNSB+varparamEigenGraf2}, and
Eq.~(\ref{StatoQNSB2}).
When $s$ increases, the GQNSB wavefunctions shift almost rigidly
toward the dissociation region of the potential. As the GQNSB
wavefunctions decay much more rapidly for small $x$ than for large
$x$, approaching $s_{\rm min}$ from below the accuracy of the
representation of the bound states localized in the well region
improves slowly, but soon after the optimal $s$ is found, all
wavefunctions move their localization region to the right of the
equilibrium position, and cease to account well for the eigenstates
behavior at the left of $x_0$.  On the other hand, $\sigma$ affects
mainly the vanishing rate for large $x$, which affects the bound
states representation quite severely, but in a rather symmetric way.
Convergency can be quite substantially improved by tuning the
wavefunctions localization, and this can be achieved by choosing the
most appropriate $s$ and $\sigma$, thus precisely the OGQNSB.

\section{Examples of Applications}
\label{AppSec}
\subsection{Ar$_2$}
\label{Ar2SecA}

We compare the OGQNSB and the QNSB for the calculation of the
vibrational spectrum of the Argon dimer, for which a reliable {\it
ab-initio} molecular potential is provided \cite{Patkowski05} in terms
of a set of 47 points in the range $x=0.25$ to $20$ \AA.  Patkowski \emph{et
al.}~\cite{Patkowski05} propose an analytic expression fitting the {\it
ab-initio} points rather accurately.  
\rem{%Tab1
\begin{table}
\caption{Fit quality and parameters for model potential
(\ref{Morse-Exp2}), $N_{\rm max}=8$, to the Ar$_2$
potential. \label{FitAr2Param}}
\vskip 0.1cm
\centerline{
\begin{tabular}{lr}
\hline
$\delta_{RMS}$  [Ha]  &  0.26   \\
$\delta_{RMS}$ well [cm$^{-1}$] & 0.48 \\
\hline
$\alpha$ & 0.516787 $a_0^{-1}$\\
$a_2$ & 1359.70868  $\mu$Ha       \\
$a_3$ & 1136.96625  $\mu$Ha       \\
$a_4$ & 181.96578   $\mu$Ha       \\
$a_5$ & 43.51541    $\mu$Ha       \\
$a_6$ & 3.77230     $\mu$Ha       \\
$a_7$ & -0.13914    $\mu$Ha       \\
$a_8$ & 0.00202     $\mu$Ha       \\
$x_0$ & 7.116       $a_0$ \\
\hline
\end{tabular}
}
\end{table}
}% Fine rem tab1
\rem{% Tab 2
\begin{table}
\caption{Energy differences (in cm$^{-1}$), between consecutive $J=0$ vibrational levels of Ar$_2$.% Funz Globale, Pesi1suEnerMan {\Large SPARISCE...}
\label{Validating}}
\vskip 0.1cm
\begin{minipage}{\textwidth}
\centerline{
\begin{tabular}{c|r|r|r}
\hline 
$i-i'$ & numerical~\footnote{Numerical diagonalization of the
Patkowski \emph{et al.}'s potential~\cite{Patkowski05}.} & OGQNSB 
& experiment\footnote{Ultraviolet laser spectroscopy data by Herman et
al .~\cite{Herman88}.}\\ 
 &  & $N_s=100$
& \\ 
\hline
1-0 & 25.76 & 25.64 & 25.69 \\
2-1 & 20.49 & 20.43 & 20.58 \\
3-2 & 15.44 & 15.46 &  15.58 \\
4-3 & 10.79 & 10.90 & 10.91 \\
5-4 & 6.75 & 6.92 & 6.84 \\
6-5 & 3.56 & 3.71 &  --\\
7-6 & 1.36 & 1.31 &  --\\
\hline
\end{tabular}}
\end{minipage}
\end{table}
} %Fine rem tab 2
We fit the {\it ab-initio} data instead to the expansion of
Eq.~(\ref{Morse-Exp2}), up to degree $N_{\rm max}=8$.
The resulting best-fit coefficients are reported in
 Table~\ref{FitAr2Param}. 
 %dissociation energy) $D_e=\sum_{i=2}^{N_{\rm max}} (-)^i a_i=99.23\,{\rm cm}^{-1}$.  
Since the repulsive small-$x$ region does not affect the
bound states significantly anyway, we privilege the convergence
inside the binding well region, with a weighted fit~\cite{Nota4}.
  Despite its simplicity, generality, and the relatively small number
of parameters involved ($N_{\rm max}+1=9$), the resulting expansion is
quite accurate, throughout the whole energy range covered by the 47
{\it ab-initio} points.  In particular, in the well region the
agreement is quite good, with a RMS discrepancy $\delta_{RMS}$ of less
than half wavenumber, see Table~\ref{FitAr2Param}.
 Moreover, the resulting model potential does not suffer from the
 unphysical small-$x$ divergence to $-\infty$ of the fitted
 function~\cite{Patkowski05}, and rather tracks the repulsive region
 within few electronvolts.  The well depth (classical dissociation
 energy) is $D_e=\sum_{i=2}^{N_{\rm max}} (-)^i a_i=99.23\,{\rm cm}^{-1}$.

 We apply the algebraic method and solve the resulting
 quantum-mechanical problem (\ref{Ham_exp2}) for the bound-state
 eigenvalues, using QNSB and GQNSB of different size $N_s$.
Table~\ref{Validating} compares the results obtained by
 finite-differences solution of the Schr\"{o}dinger equation for the
 analytic potential by Patkowski \emph{et al.}~\cite{Patkowski05}, and
 by numerical diagonalization of the algebraic Hamiltonian
 (\ref{Ham_exp2}) with the parameters from Table~\ref{FitAr2Param} on
 a large $N_s=100$ OGQNSB ($s=80.18$, $\sigma=0.213$).
This large OGQNSB was chosen to ensure that the results are fully
converged, and is taken as reference.  
The excitation energies obtained using our expansion compare
favourably to those obtained by using Patkowski \emph{et al.} analytic
expression\cite{Patkowski05}, and to the experimental $J=0$
data~\cite{Herman88}, demonstrating equally good or better agreement.

Table~\ref{EnerAr2AllAll} illustrates the convergency properties
of the unoptimized QNSB by reporting the eigenvalues obtained by
diagonalizing the expanded Hamiltonian (\ref{Ham_exp2}) on $N_s=100$,
$20$ and $15$ states respectively.  The energy differences with
respect to the $N_s=100$ OGQNSB reference are shown, in parentheses,
when exceeding $10^{-3}$ cm$^{-1}$.  Fairly well converged results are
obtained even for the small $N_s=15$ QNSB. Notice however that the
bound state closest to dissociation is unbound for $N_s=15$ and $20$,
since it is so extended that a rather large QNSB ($N_s\geq 42$) is
needed to obtain it at negative energy. Even the very large $N_s=100$
QNSB does not provide a well-converged result for that specific level.

By diagonalizing the expanded Hamiltonian (\ref{Ham_exp2}) on $N_s=20$
and $N_s=15$ OGQNSB, we obtain the complete spectrum, and with an
accuracy $\Delta-\Delta_0$ of $6\cdot 10^{-5}$ and $0.013$ cm$^{-1}$
respectively. The accuracy of all bound levels but the last one is
basically the same as for the corresponding QNSB, but the complete
discrete spectrum is obtained, including the highest level.
The accuracy of the $N_s=15$ OGQNSB is therefore better than that of
$N_s=100$ QNSB, for Ar$_2$. Reducing the basis size below $N_s=15$,
the highest state is missing, but the GQNSB can still be tuned to
obtain a fair accuracy of all other states ($\Delta-\Delta_0<0.5~ {\rm
cm}^{-1}$ for $N_s\geq 11$).

\rem{ %Tab 3
\begin{table}
\caption{Bound-state eigenvalues of the Ar$_2$ dimer, computed with different methods, all based on the {\it ab-initio} values~\cite{Patkowski05} [in cm$^{-1}$].  \label{EnerAr2AllAll}}
\vskip 0.1 cm
\begin{minipage}{\textwidth}
\centerline{
\begin{tabular}{l|c|c|c|c}
\hline
state & OGQNSB & QNSB     &  QNSB      & QNSB\\
 &      $N_s=100$  & $N_s=100$&  $N_s=20$  & $N_s=15$ \\
\hline
0 & -84.41     & -84.41 (-)    & -84.41 (-)  & -84.40 (0.005) \\ 
1 & -58.77     & -58.77 (-)    & -58.77 (-)  & -58.75 (0.01) \\ 
2 & -38.34     & -38.34 (-)    & -38.34 (-)  & -38.31 (0.02) \\ 
3 & -22.88     & -22.88 (-)    & -22.88 (-)  & -22.84 (0.04) \\ 
4 & -11.98     & -11.98 (-)    & -11.98 (-)  & -11.93 (0.05) \\ 
5 & -5.06      & -5.06  (-)    & -5.06 (-)   & -5.02 (0.04) \\ 
6 & -1.35      & -1.35  (-)    & -1.35 (-)   & -1.32 (0.02) \\ 
7 & -0.036     & -0.015 (0.02) & {\it 0.022 (0.06)}         & {\it 0.051 (0.09)} \\
\hline
\end{tabular}}
\end{minipage}
\end{table}
}%Fine rem Tab 3

\subsection{H$_2$}
\label{H2Sec}

In a previuos work~\cite{Bordoni_2006} we applied the QNSB formalism
to the {\it ab-initio} adiabatic potential~\cite{SLR87,KolWol2} for
the H$_2$ molecule.
We found that an expansion (\ref{Morse-Exp2}) up to $N_{\rm max}=12$
fits all 169 available {\it ab-initio} points with a deviation
$\delta_{RMS}=5.5$ cm$^{-1}$. This expansion, whose parameters are
reported in Table~IV of Ref.~\cite{Bordoni_2006}\,\!, produces all the 15
vibrational bound states of this molecule. The QNSB parameters for
this potential are $ s=25.56$ and $\sigma=0.564$. The QNSB
produces a cm$^{-1}$ converged spectrum using $N_s\geq 28$ basis
states.

By minimizing $\tilde{\Delta}$, Eq.~(\ref{QNSB+Minim2}), we generate
an OGQNSB of smaller $N_s$.  For the calculation of $\tilde{\Delta}$ we
use the fully converged $N_s=200$ QNSB results as reference, reported
in the second column of Table~\ref{EigH2-21}.
A $N_s=25$ OGQNSB with $s=26.36$ and $\sigma=2.115$
($\tilde{\Delta}=0.173$ cm$^{-1}$) produces eigenvalues with the same
cm$^{-1}$ figures, i.e. the same accuracy of the $N_s=28$ QNSB: since
they are identical to the second column of Table \ref{EigH2-21}, they
are not shown. For less strict accuracy requirements, one could reduce
the basis size: the last two columns of Table~\ref{EigH2-21} compare the
eigenvalues obtained with $N_s=21$ QNSB and OGQNSB.
The H$_2$ potential expansion illustrates the robustness of the GQNSB
in state-poor situations: 
here, for the $N_s=21$ QNSB eigenvalues the differences with respect
to the fully converged values reach hundreds of wavenumbers, with a RMS
discrepancy $\tilde{\Delta}=323$ cm$^{-1}$, while the discrepancy of
the eigenvalues obtained by diagonalizing on the $N_s=21$ OGQNSB
amounts to $\tilde{\Delta}=3.7$ cm$^{-1}$ only.

\rem{ % Table 4
\begin{table}
\caption{\label{EigH2-21}
H$_2$ bound-state energies in reduced-size algebraic bases; in parentheses, the differences w.r.t. the reference [in cm$^{-1}$].%\footnote{
}
\vskip 0.1 cm
\begin{minipage}{\textwidth}
\centerline{
\begin{tabular}{c|l|l|l}%\label{EigH2}                      60                        40
\hline
state &  QNSB\footnote{Reference fully converged calculation.} & QNSB\footnote{
Eigenvalues obtained with a $N_s=21$ QNSB. The maximum
difference of 596 cm$^{-1}$ corresponds to
1.6\% of the well depth.
} & OGQNSB\footnote{
Eigenvalues obtained with a $N_s=21$ OGQNSB ($s=23.52$, and $\sigma=3.194$). The maximum
difference of 10 cm$^{-1}$ corresponds to
0.03\% of the well depth.
}\\
  &   $N_s=200$       &   $N_s=21$   &   $N_s=21$     \\
\hline 
0 & -36113  & -36113 (-) & -36113 (-) \\
1 & -31948  & -31948 (-) & -31948 (-) \\
2 & -28020  & -28019 (1) & -28020 (-) \\
3 & -24324  & -24322 (2) & -24324 (-) \\
4 & -20856  & -20845 (11) & -20856 (-) \\
5 & -17614  & -17573 (41) & -17614 (-) \\
6 & -14599  & -14486 (113) & -14598 (1) \\
7 & -11815  & -11588 (227) & -11814 (2) \\
8 & -9271   & -8907 (365) & -9269 (3) \\
9 & -6979   & -6486 (493) & -6975 (4) \\
10 & -4955  & -4376 (579) & -4951 (4) \\
11 & -3222  & -2626 (596) & -3218 (4) \\ 
12 & -1810  & -1281 (530) & -1806 (5) \\
13 & -761   & -389 (372) & -757 (4) \\
14 & -135   & -9 (126) & -125 (10) \\
\hline
\end{tabular}}
\end{minipage} 
\end{table} 
} % Fine rem Tab 4

\section{Conclusions}
\label{ConcSect}

The substantial improvement of the variational accuracy of the
bound-state spectra computed on a OGQNSB w.r.t. the unoptimized QNSB
permits in practice to make calculations of a given accuracy on a
significantly smaller basis size. While this improvement is
practically irrelevant to the solution of the $1$-dimensional
vibrational problem of diatomics, it is of great importance for the
application of this method to the calculation of the spectra based on
the {\it ab-initio} multi-dimensional potential surfaces of polyatomic
molecules, as is currently pursued in quantum chemical
research~\cite{Wyatt98,Pochert00,Callegari03,Handy04,Zobov07,Makarewicz07}.
We are currently testing the generalization of the
expansion~(\ref{Morse-Exp2}) to the polyatomic
case\cite{Bordoni2006_T}.

\section*{Acknowledgement}
\label{AcknSect} 
 
We thank Konrad Patkowski for kindly providing us with the complete
{\it ab-initio} Ar$_2$ PES data including those not available in his
paper~\cite{Patkowski05}.

\newpage

\begin{table}[h]
\caption{Fit quality and parameters for model potential
(\ref{Morse-Exp2}), $N_{\rm max}=8$, to the Ar$_2$
potential. \label{FitAr2Param}}
\vskip 0.1cm
\centerline{
\begin{tabular}{lr}
\hline
$\delta_{RMS}$  [Ha]  &  0.26   \\
$\delta_{RMS}$ well [cm$^{-1}$] & 0.48 \\
\hline
$\alpha$ & 0.516787 $a_0^{-1}$\\
$a_2$ & 1359.70868  $\mu$Ha       \\
$a_3$ & 1136.96625  $\mu$Ha       \\
$a_4$ & 181.96578   $\mu$Ha       \\
$a_5$ & 43.51541    $\mu$Ha       \\
$a_6$ & 3.77230     $\mu$Ha       \\
$a_7$ & -0.13914    $\mu$Ha       \\
$a_8$ & 0.00202     $\mu$Ha       \\
$x_0$ & 7.116       $a_0$ \\
\hline
\end{tabular}
}
\end{table}

\begin{table}
\caption{Energy differences (in cm$^{-1}$), between consecutive $J=0$ vibrational levels of Ar$_2$.% Funz Globale, Pesi1suEnerMan {\Large SPARISCE...}
\label{Validating}}
\vskip 0.1cm
\begin{minipage}{\textwidth}
\centerline{
\begin{tabular}{c|r|r|r}
\hline 
$i-i'$ & numerical~\footnote{Numerical diagonalization of the
Patkowski \emph{et al.}'s potential~\cite{Patkowski05}.} & OGQNSB 
& experiment\footnote{Ultraviolet laser spectroscopy data by Herman et
al .~\cite{Herman88}.}\\ 
 &  & $N_s=100$
& \\ 
\hline
1-0 & 25.76 & 25.64 & 25.69 \\
2-1 & 20.49 & 20.43 & 20.58 \\
3-2 & 15.44 & 15.46 &  15.58 \\
4-3 & 10.79 & 10.90 & 10.91 \\
5-4 & 6.75 & 6.92 & 6.84 \\
6-5 & 3.56 & 3.71 &  --\\
7-6 & 1.36 & 1.31 &  --\\
\hline
\end{tabular}}
\end{minipage}
\end{table}

\begin{table}
\caption{Bound-state eigenvalues of the Ar$_2$ dimer, computed with different methods, all based on the {\it ab-initio} values~\cite{Patkowski05} [in cm$^{-1}$].  \label{EnerAr2AllAll}}
\vskip 0.1 cm
\begin{minipage}{\textwidth}
\centerline{
\begin{tabular}{l|c|c|c|c}
\hline
state & OGQNSB & QNSB     &  QNSB      & QNSB\\
 &      $N_s=100$  & $N_s=100$&  $N_s=20$  & $N_s=15$ \\
\hline
0 & -84.41     & -84.41 (-)    & -84.41 (-)  & -84.40 (0.005) \\ 
1 & -58.77     & -58.77 (-)    & -58.77 (-)  & -58.75 (0.01) \\ 
2 & -38.34     & -38.34 (-)    & -38.34 (-)  & -38.31 (0.02) \\ 
3 & -22.88     & -22.88 (-)    & -22.88 (-)  & -22.84 (0.04) \\ 
4 & -11.98     & -11.98 (-)    & -11.98 (-)  & -11.93 (0.05) \\ 
5 & -5.06      & -5.06  (-)    & -5.06 (-)   & -5.02 (0.04) \\ 
6 & -1.35      & -1.35  (-)    & -1.35 (-)   & -1.32 (0.02) \\ 
7 & -0.036     & -0.015 (0.02) & {\it 0.022 (0.06)}         & {\it 0.051 (0.09)} \\
\hline
\end{tabular}}
\end{minipage}
\end{table}

\begin{table}
\caption{\label{EigH2-21}
H$_2$ bound-state energies in reduced-size algebraic bases; in parentheses, the differences w.r.t. the reference [in cm$^{-1}$].%\footnote{
}
\vskip 0.1 cm
\begin{minipage}{\textwidth}
\centerline{
\begin{tabular}{c|l|l|l}%\label{EigH2}                      60                        40
\hline
state &  QNSB\footnote{Reference fully converged calculation.} & QNSB\footnote{
Eigenvalues obtained with a $N_s=21$ QNSB. The maximum
difference of 596 cm$^{-1}$ corresponds to
1.6\% of the well depth.
} & OGQNSB\footnote{
Eigenvalues obtained with a $N_s=21$ OGQNSB ($s=23.52$, and $\sigma=3.194$). The maximum
difference of 10 cm$^{-1}$ corresponds to
0.03\% of the well depth.
}\\
  &   $N_s=200$       &   $N_s=21$   &   $N_s=21$     \\
\hline 
0 & -36113  & -36113 (-) & -36113 (-) \\
1 & -31948  & -31948 (-) & -31948 (-) \\
2 & -28020  & -28019 (1) & -28020 (-) \\
3 & -24324  & -24322 (2) & -24324 (-) \\
4 & -20856  & -20845 (11) & -20856 (-) \\
5 & -17614  & -17573 (41) & -17614 (-) \\
6 & -14599  & -14486 (113) & -14598 (1) \\
7 & -11815  & -11588 (227) & -11814 (2) \\
8 & -9271   & -8907 (365) & -9269 (3) \\
9 & -6979   & -6486 (493) & -6975 (4) \\
10 & -4955  & -4376 (579) & -4951 (4) \\
11 & -3222  & -2626 (596) & -3218 (4) \\ 
12 & -1810  & -1281 (530) & -1806 (5) \\
13 & -761   & -389 (372) & -757 (4) \\
14 & -135   & -9 (126) & -125 (10) \\
\hline
\end{tabular}}
\end{minipage} 
\end{table} 

\newpage

\begin{figure}%[h]
\caption{QNSB  wavefunctions for $n=0$, $n=4$ and $n=8$, compatible with a Morse problem characterized by $x_0=1,\alpha=4/x_0,a_2=625$, in units where $\mu=1, \hbar=1$, so that $s=8.34, \sigma=0.34$.  \label{QNSBEigenGraf2}}
\end{figure}     

\begin{figure}  
\caption{Variation of the $n=4$ GQNSB wavefunction for a $50\%$ increase in the parameters $s$ (a), or $\sigma$ (b), solid line, with respect to the QNSB starting wavefunction (dashed line) corresponding to $s=8.34$, $\sigma=0.34$, $x_0=1$, $\alpha=4/x_0$, like in Fig.~\ref{QNSBEigenGraf2}.  \label{QNSB+varparamEigenGraf2}}  
\end{figure} 

\begin{figure}%\centerline{
\caption{Discrepancies $(E_n-E^{\rm ex}_n)/a_2$ of the individual
eigenvalues $n$ for the potential $V(x)= a_2 [(v(x))^{2} +
(v(x))^{4}]$. (a) $N_s=30$ (OGQNSB with $s_{\rm min}=20.01$,
$\sigma_{\rm min}=0.435$); (b) $N_s=16$ (OGQNSB with $s=14.47$,
$\sigma=0.314$), for all bound states. The OGQNSB (diamonds)
discrepancies are compared to those based on the QNSB (circles), and
to the GQNSB based on the choice of $(s,\sigma)$ made by Tennyson and
Sutcliffe \cite{Tennyson82,Tennyson04} (triangles).\label{DiffComp2} }
\end{figure} 

\begin{figure}
\caption{$s$ dependence of $\Delta$ Eq.~(\ref{QNSB+MinimEig2}), for $V(x)$ defined in Eq.~(\ref{ToyModel}), computed with the GQNSB of $N_s=30$ elements, as a function of $s$, and for $\sigma$ fixed to $\sigma_{\rm min}$, to $\sigma_{\rm min} \pm0.005,$ and to $\sigma_{\rm min} \pm 0.01$. \label{sandsigmadependence2}}
\end{figure}

\newpage

\setcounter{figure}{0}

\begin{figure}[ht]
\centerline{ 
\begin{tabular}{c}
\epsfig{file=xQNSBCombD.eps,width=5.5cm,clip=}\\
\end{tabular}}\caption{%QNSB  wavefunctions, for $n=0$, $n=4$ and $n=8$, compatible with a Morse problem characterized by $x_0=1,\alpha=4/x_0,a_2=625$, in units where $\mu=1, \hbar=1$, so that $s=8.34, \sigma=0.34$.  \label{QNSBEigenGraf2}
}
\end{figure}     

\begin{figure}  
\centerline{\begin{tabular}{c}
\epsfig{file=xQNSBMidCombD20.5b.eps,width=10.0cm,clip=}\\
\end{tabular}}\caption{%Variation of the $n=4$ GQNSB wavefunction for a $50\%$ increase in the parameters $s$ (a), or $\sigma$ (b), solid line, with respect to the QNSB starting wavefunction (dashed line) corresponding to $s=8.34$, $\sigma=0.34$, $x_0=1$, $\alpha=4/x_0$ like in Fig.~\ref{QNSBEigenGraf2}.  \label{QNSB+varparamEigenGraf2}
}  
\end{figure} 

\begin{figure}\centerline{
\begin{tabular}{c} 
\epsfig{file=Quart1_30_16scart2.eps,width=12cm,clip=}
\end{tabular}} 
\caption{%Discrepancies $(E_n-E^{\rm ex}_n)/a_2$ of the individual eigenvalues $n$ for the potential $V(x)= a_2 [(v(x))^{2} + (v(x))^{4}]$. (a) $N_s=30$ (OGQNSB with $s_{\rm min}=20.01$, $\sigma_{\rm min}=0.435$); (b) $N_s=16$ (OGQNSB with $s=14.47$, $\sigma=0.314$), for all bound states. The OGQNSB (diamonds) discrepancies are compared to those based on the QNSB (circles), and to the GQNSB based on the choice of $(s,\sigma)$ made by Tennyson and Sutcliffe \cite{Tennyson82,Tennyson04} (triangles).\label{DiffComp2} 
}
\end{figure} 

\begin{figure}
\centerline{
\begin{tabular}{c}
\epsfig{file=EsseperdeltasigI2newb.eps,width=7cm,clip=}\\
\end{tabular}}\caption{%$s$ dependence of $\Delta$ Eq.~(\ref{QNSB+MinimEig2}), for $V(x)$ defined in Eq.~(\ref{ToyModel}), computed with the GQNSB of $N_s=30$ elements, as a function of $s$, and for $\sigma$ fixed to $\sigma_{\rm min}$, to $\sigma_{\rm min} \pm0.005,$ and to $\sigma_{\rm min} \pm 0.01$. \label{sandsigmadependence2}
}
\end{figure}


\begin{thebibliography}{21}
\bibitem{Nota1} {One could even consider different $q$ for the $A$ and
$A^{\dagger}$ operators, but such more general choice does not seem
particularly interesting for molecular applications: we take a single
$q$.}
\bibitem{Nota2} {The use of an $\alpha$ basis-parameter different from
the potential one affects the matrix elements substantially. This is
seen through the recursive relations for the matrix
elements~\cite{Ben-Mol-Al}, specialized to the present case (subscript
$B$ and $P$ stand for basis and potential respectively)
\begin{eqnarray}\label{MatelGen12}\nonumber
\langle \phi_m|e^{-\frac{\alpha_P}{\alpha_B} \alpha_B (\hat{x} -x_0)}|\phi_{n+1} \rangle&=&[C_m \langle \phi_{m-1}|e^{-\frac{\alpha_P}{\alpha_B} \alpha_B (\hat{x} -x_0)}|\phi_n \rangle + \\ \nonumber
&&(n-m-\frac{\alpha_P}{\alpha_B}) \langle \phi_{m}|e^{-\frac{\alpha_P}{\alpha_B} \alpha_B (\hat{x} -x_0)}|\phi_n \rangle ]/C_{n+1}\,,\\\nonumber
\langle \phi_{0}|e^{-\frac{\alpha_P}{\alpha_B} \alpha_B (\hat{x} -x_0)}|\phi_0 \rangle &=&\frac{\Gamma(2 \sigma + \frac{\alpha_P}{\alpha_B})}{(2s + 1 )^{\frac{\alpha_P}{\alpha_B}} \Gamma(2 \sigma)}\,,
\end{eqnarray}
where the $C_{n}$ are the coefficient of Eq.~(\ref{CCdef2}). 
This recursion implies that if $\alpha_P \neq k
\alpha_B$, for integer $k$, all matrix elements of integer powers of
$e^{-\alpha_P (\hat{x}-x_{0})}$ are non-vanishing.  Moreover, these
recursive relations are more intricate than the
algebraic rules holding when $\alpha_P/\alpha_B$ is an integer.}
\bibitem{Nota3} {In addition, Eqs.~(\ref{SigFix2}) and (\ref{SFix2})
grant an isomorphism of the Morse-problem bound-states subspace with
the subspace generated by the first $[s]+1$ states of the QNSB
\cite{Ben-Mol-Al}; this isomorphism does not hold for general $s$ and
$\sigma$.}
\end{thebibliography}
\end{document}